\documentclass{jaa}
\usepackage[round]{natbib}
\usepackage{newtxtext,newtxmath}
\usepackage[T1]{fontenc}
\usepackage{amsmath}

\usepackage{float}
\usepackage{graphicx}

\usepackage{geometry}
\usepackage{array}
\usepackage{orcidlink}
\usepackage{supertabular,booktabs}
\usepackage{booktabs}
\usepackage{scalerel}
\usepackage{placeins}
\usepackage{adjustbox}
\usepackage{hyperref,url}
\hypersetup{pdfauthor={Name}}
\usepackage[online]{threeparttablex}
\usepackage{siunitx}

\hypersetup{%
colorlinks=true,linkcolor=blue,anchorcolor=black,citecolor=blue,filecolor=black,menucolor=black,runcolor=black,urlcolor=blue
}
\newcounter{bibcount}

\makeatletter
\patchcmd{\@lbibitem}{\item[}{\item[\hfil\stepcounter{bibcount}{\thebibcount.}}{}{}
\renewcommand\NAT@bibsetup%
   [1]{\setlength{\leftmargin}{\bibhang}\setlength{\itemindent}{-\parindent}%
       \setlength{\itemsep}{\bibsep}\setlength{\parsep}{\z@}}
\makeatother
\DeclareUnicodeCharacter{2212}{-}
\usepackage{titlesec}

\usepackage{graphicx}

\begin{document}\sloppy
\title{Possibilities of Identifying Members from Milky Way Satellite Galaxies using Unsupervised Machine Learning Algorithms}

\author{Devika K Divakar \orcidlink{0000-0001-8428-1222}\textsuperscript{1,2,*},  Pallavi Saraf\orcidlink{0009-0001-4813-0432}\textsuperscript{1,3}, Thirupathi Sivarani \orcidlink{0000-0003-0891-8994}\textsuperscript{1}, Vijayakumar H Doddamani\orcidlink{0009-0008-3515-257X}\textsuperscript{2}}
\affilOne{\textsuperscript{1}Indian Institute of Astrophysics, Bangalore - 560034, India\\}
\affilTwo{\textsuperscript{2}Department of Physics, Bangalore University, Bangalore - 560056, India\\}
\affilThree{\textsuperscript{3}Pondicherry University, R.V. Nagar, Kalapet, 605014, Puducherry, India}

\twocolumn[{

\maketitle

\corres{devika.itcc@iiap.res.in}

\msinfo{}{}

\begin{abstract}
A detailed study of stellar populations in Milky Way (MW) satellite galaxies remains an observational challenge due to their faintness and fewer spectroscopically confirmed member stars. We use unsupervised machine learning methods to identify new members for nine nearby MW satellite galaxies using Gaia data release-3 (Gaia DR3) astrometry and  the Dark Energy Survey (DES) and the DECam Local Volume Exploration Survey (DELVE)  photometry. Two density-based clustering algorithms, DBSCAN and HDBSCAN, have been used in the four-dimensional astrometric parameter space ({\fontfamily{lmtt}\selectfont $\alpha_{2016}$, $\delta_{2016}$, $\mu_{\alpha} cos\delta$, $\mu_\delta $}) to identify member stars belonging to MW satellite galaxies. Our results indicate that we can recover more than 80\% of the known spectroscopically confirmed members in most of the satellite galaxies and also reject 95-100\% of spectroscopic non-members. We have also added many new members using this method.  We compare our results with previous studies that also use photometric and astrometric data and discuss the suitability of density-based clustering methods for MW satellite galaxies
\end{abstract}

}]

\doinum{12.3456/s78910-011-012-3}
\artcitid{\#\#\#\#}
\volnum{000}
\pubyear{2023}
\pgrange{1--13}
\setcounter{page}{1}
\lp{20}
\vspace{2cm}
\section{Introduction}
Dwarf satellite galaxies play a fundamental role in the galaxy evolution paradigm. They are believed to be the first galaxies resulting from the hierarchical merging of substructures arising from primordial density fluctuations, eventually forming massive galaxies like our MW \citep{Kravtsov_1998, Bullock_2005}. Therefore, MW satellite galaxies serve as exceptional opportunities to investigate galaxy formation at close proximity \citep{Rey_2019}. The study of these satellite galaxies is essential for understanding the early chemical evolution of the MW \citep{Frebel_2012, Frebel_2014} and provides indirect evidence of interactions with dark matter \citep{Abdallah_2020,Acciari_2022}. Additionally, MW satellite galaxies serve as unique opportunities to investigate the theories of dark matter, for example, the "Missing Satellites Problem", that is, the overabundance of substructures such as MW satellites that are expected in Cold Dark Matter (CDM) paradigm, and the known deficit in the number of MW satellites from observations \citep{Kauffmann_1993,Klypin_1999,Moore_1999}. 
However, the discovery of a vast population of new, ultra-faint dwarf(UFD) galaxies in Local Group as a result of wide-field, resolved star surveys (Sloan
Digital Sky Survey (SDSS); \citep{York_2000}, Dark Energy Survey
(DES); \citep{Diehl_2014}, Pan-STARRS; \citep{Chambers_2016}, Gaia; \citep{Gaia-Collaboration_2018}) that have more than doubled the dwarf satellite count in the recent years \citep{Drlica-Wagner_2015,Laevens_2015a,Laevens_2015b,Koposov2015,Bechtol_2015,Drlica-Wagner_2016, Kim_Jerjen_2015, Kim_2015, Homma_2016, Homma_Chiba_2018, Torrealba, Mau_2020, Cerny_2021, Cerny_2023}, along with spectroscopic follow up studies by \citep{Simon, Kirby_2013, Weisz_2015, Ji_2016,Koposov, Li2018b, Ji_2020, Jenkins_2021, Waller_2022}.While many of the largest and most prominent satellite galaxies have been well-studied, cold dark matter simulations project that there are likely to be many smaller and more distant satellite galaxies that remain undiscovered due to observational constraints, including in areas surveyed by previous sky surveys \citep{Koposov_2008, Tollerud_2008, Hargis_2014, Newton_2018, Nadler_2020, Manwadkar_2022}. The revised perspective challenges the notion of the missing satellites problem, suggesting that the observed population of satellite galaxies around the MW is consistent with theoretical predictions once observational biases and limitations are adequately accounted for \citep{Kim_2018}.\par The current Gaia mission's data release-3 offers accurate sub-milli-arcsecond astrometry and homogeneous three-band photometry for over 1.3 billion stars \citep{Gaiadr3_2022}. Using machine learning algorithms and other automatic clustering strategies becomes vital for uncovering patterns and classifying objects in such a big astronomical data frame (\citep{Ball_2010, Baron_2019}; references therein). In contrast to supervised learning, where the algorithm undergoes training on labeled data to predict or categorize new instances \citep{Odewahn_1992, Weir_1995, Vasconcellos_2011}, unsupervised learning algorithms work on unstructured data with no pre-existing labels or targets. This enables them to uncover previously unidentified patterns and relationships in data, providing novel insights and discoveries \citep{Rubin_2016, Baron_2017, Elvin-Poole_2018, Reis_2021, Nidever_2021}. This paper presents the possibility of identifying new members of known nearby {MW satellite galaxies} using unsupervised clustering algorithms.  

In Section \ref{sec-data}, we discuss the {MW satellite galaxies} that are considered in this study and a summary of the astrometric Gaia DR3, the supplementary photometric data sets, and our preliminary quality criteria. In Section \ref{sec-methods}, we explain the unsupervised learning algorithms we used in this paper and additional color-magnitude selection criteria, along with an illustrative example showcasing the validation of hyperparameter selection through a comparison of our results with spectroscopically confirmed members. In Section \ref{sec-results-discussions}, we present the results and a comprehensive discussion on the comparison of unsupervised learning algorithms employed in this study. We summarize and conclude in Section \ref{sec-conclusion}

\section{Target {satellite galaxies} and Data}
\label{sec-data}
 {MW satellite galaxies} that are nearby and have extensive spectroscopic follow-up have been selected in this study. Spectroscopic results will help understand the efficiency of the density-based clustering algorithms and derive optimized hyperparameters. Most targets are within 50 kilo-parsec (kpc) distance, and that is observed by the {DECam \citep{Flaugher_2015}}. Some of the targets (e.g., BoöI) are at a larger distance, but they have more than several hundred spectroscopically confirmed members and a well-populated Red Giant Branch (RGB) and Horizontal Branch (HB) stars (compared to other UFDs) and hence selected for the study.
\subsection{Astrometric Data}
Astrometric data from the Gaia DR3 archive \citep{Gaiadr3_2022} over a field of view of 1\textdegree $\times$ 1\textdegree \: around the center of each satellite is considered for most of the targets, which is much larger than the half-light radius of a typical nearby UFD. Nevertheless, {for galaxies with extended structure (CarII, BoöI, HyiI, TucIII), we have considered a larger field of view (with a diameter 1.5\textdegree – 7\textdegree \:)  to cover the spectroscopic members.} Table \ref{table:table1} lists the known structural parameters and extinction in the direction of the targets. { In each of the target fields, we utilize the complete dataset from Gaia DR3 for sources that have recorded measurements of position and proper motion (PM)({\fontfamily{lmtt}\selectfont $\alpha_{2016}$, $\delta_{2016}$, $\mu_{\alpha} cos\delta$, $\mu_\delta $}). The high-quality astrometry as defined by the renormalized unit-weighted error ({\fontfamily{lmtt}\selectfont ruwe} < 1.4; \citet{Fabricius_2021}) and {\fontfamily{lmtt}\selectfont astrometric\_excess\_noise\_sig} <= 2; \citet{Lindegren_2021} has been used to select the targets. The parallax cut ($\varpi$ - 3$\sigma_{\varpi}$ < 0.0mas) is used to remove the foreground stars.}
\subsection{Photometric Data}
Most of the nearby UFDs have main sequence turnoff magnitude around Vmag $= 21-22$, and Gaia photometry has significant errors at fainter magnitudes $(G \geq 20)$, color-magnitude selection of the astrometrically pre-selected sample is hence not accurate. Employing precise photometric measurements yields a marked reduction in foreground contamination, thus prompting us to incorporate more accurate photometry from the Dark Energy Survey (DES) DR2
\citep{Abbott_2021} and DECam Local Volume Exploration Survey (DELVE) {DR2 \citep{Drlica-Wagner_2022}.}  Figure \ref{fig1:cmd-des-gaia} shows the color-magnitude diagram (CMD) of all the spectroscopic members in Gaia and DECam magnitudes, along with a 12.5 Gyr isochrone (details in Section \ref{subsec-isosel}). For Gaia CMD (\textit{Top} panel), the colorbar shows the error in BP magnitude. For Gaia photomerty, we used ${A}_{G}\,=2.740\ *E(B-V)$ \citep{Casagrande_2018} for the reddening correction. The excess color $E(B − V)$ \citep{Schlegel_1998} was acquired using the python task {\fontfamily{lmtt}\selectfont dustmaps} \citep{Green_2018}. 
The \textit{Bottom} panel in Figure \ref{fig1:cmd-des-gaia} shows the DECam CMD of spectroscopic members and the error in $r$ magnitude. For DES photometry, dereddened magnitudes (weighted-average psf magnitudes $wavg\_mag\_psf\_\{g,r\}\_dered$) have been used, which were derived using $E(B − V)$ values from \cite{Schlegel_1998}. {In the case of DELVE photometry, we used weighted-average psf magnitudes ($wavg\_mag\_psf\_\{g,r\}$) along with $extinction\_\{g,r\}$ for extinction correction.}
\subsection{Isochrone Selection}
\label{subsec-isosel}
\begin{figure}[ht!]
    \centering
    \includegraphics[width=0.38\textwidth]{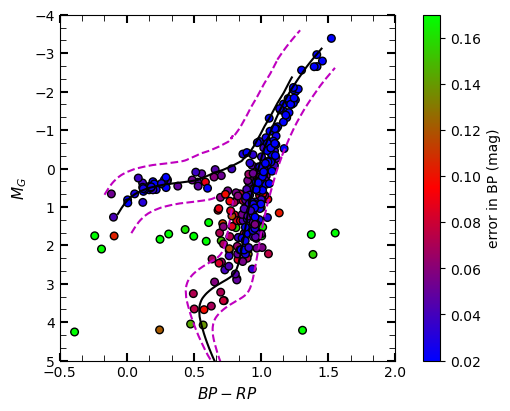}\\\includegraphics[width=0.38\textwidth]{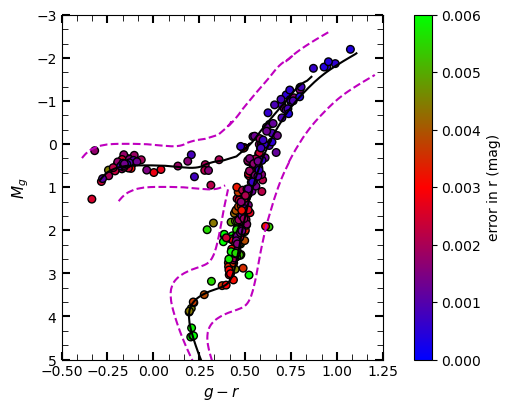}
   \caption{{\footnotesize \textit{Top}:  Color Magnitude Diagram of spectroscopic members in Gaia (References given in Table \ref{table:table1}). Colorbar shows the error in BP. Black curve is BaSTi isochrone of age 12.5Gyr and [Fe/H] = -2.2. Magenta dotted line shows the color-magnitude selection ($\Delta (G_{BP} − G_{RP})$ = ±0.1 mag) and magnitude ($\Delta G$ = ±0.5 mag). \textit{Bottom}: Same as \textit{Top} panel, but for DECam based CMD with color-magnitude selection ($\Delta (g − r)$ = ±0.1 mag) and magnitude ($\Delta g$ = ±0.5 mag).}}
    \label{fig1:cmd-des-gaia}
\end{figure}

Before implementing clustering algorithms on the sample data, we perform positional cross-match between the final satellite samples in Gaia and DECam photometry. Furthermore, we apply an additional selection criteria on the CMD using DECam photometry. The isochrone is selected based on the spectroscopically confirmed members of the {satellite galaxies}. Table \ref{table:table1} provides some features of the  {satellite galaxies} and the corresponding references. As seen in the \textit{Bottom} panel of Figure \ref{fig1:cmd-des-gaia}, most of the spectroscopically confirmed members lie on a BaSTI (Bag of Stellar Tracks and Isochrones; \citet{Pietrinferni_2021}) isochrone with an old and metal-poor population of age = 12.5 Giga-year (Gyr) and [Fe/H] = −2.2. Thus, we restrict our selection of candidate members (comprising main-sequence turnoff (MSTO), RGB, and HB  stars) to those lying in close proximity to this isochrone. We selected all the targets to be within a liberal range of color ($\Delta (g − r)$ = ±0.1 mag) and magnitude ($\Delta g$ = ±0.5 mag) to this isochrone, as illustrated by a magenta dotted line in Figure \ref{fig1:cmd-des-gaia}.
\renewcommand{\arraystretch}{1.6}
\setlength{\tabcolsep}{1.5pt}
\begin{table*}
\caption{Properties of {Milky Way Satellite galaxies}}
\vspace{-2pt}
\begin{center}
{\scriptsize
\begin{tabular}{|cccccccccccc|}
\hline
Satellite&$\alpha${\tiny (deg)} &$\delta${\tiny (deg)}&$N_{\text{spec}}$&D{\tiny (kpc)}&(m-M)&E(B$-$V)&$\epsilon$&$\theta$°&$M_{v}$&$r_{h}'$&Reference\\ 
\hline
\tablelasttail{\hline}
Tucana III & 359.15 & -59.6 & 48 & $25.00_{\scaleto{-2.00}{4pt}}^{\scaleto{+2.00}{4pt}}$ & $17.01_{\scaleto{-0.10}{4pt}}^{\scaleto{+0.10}{4pt}}$ & 0.011 & ... & ... & $-2.40_{\scaleto{-0.42}{4pt}}^{\scaleto{+0.42}{4pt}}$ & $6.00_{\scaleto{-0.6}{4pt}}^{\scaleto{+0.8}{4pt}}$ & \scriptsize{[1],RV[2,3]} \\
Hydrus I & 37.39 & -79.31 & 33 & $27.60_{\scaleto{-0.50}{4pt}}^{\scaleto{+0.50}{4pt}}$ & $17.20_{\scaleto{-0.04}{4pt}}^{\scaleto{+0.04}{4pt}}$ & 0.090 & $0.21_{\scaleto{-0.07}{4pt}}^{\scaleto{+0.15}{4pt}}$ & $97.0_{\scaleto{-14.0}{4pt}}^{\scaleto{+14.0}{4pt}}$ & $-4.71_{\scaleto{-0.08}{4pt}}^{\scaleto{+0.08}{4pt}}$ & $7.42_{\scaleto{-0.54}{4pt}}^{\scaleto{+0.62}{4pt}}$ & \scriptsize{[4],RV[4]}\\
Carina III & 114.63 & -57.89 & 5 & $27.80_{\scaleto{-0.60}{4pt}}^{\scaleto{+0.60}{4pt}}$ & $17.22_{\scaleto{-0.10}{4pt}}^{\scaleto{+0.10}{4pt}}$ & 0.195 & $0.55_{\scaleto{-0.18}{4pt}}^{\scaleto{+0.18}{4pt}}$ & $150.0_{\scaleto{-14.0}{4pt}}^{\scaleto{+14.0}{4pt}}$ & $-2.40_{\scaleto{-0.20}{4pt}}^{\scaleto{+0.20}{4pt}}$ & $3.75_{\scaleto{-1.00}{4pt}}^{\scaleto{+1.00}{4pt}}$ & \scriptsize{[5],RV[6]}\\
Reticulum II & 53.92 & -54.05 & 28 & $30.00_{\scaleto{-2.00}{4pt}}^{\scaleto{+2.00}{4pt}}$ & $17.40_{\scaleto{-0.20}{4pt}}^{\scaleto{+0.20}{4pt}}$ & 0.018 & $0.56_{\scaleto{-0.03}{4pt}}^{\scaleto{+0.03}{4pt}}$ & $69.00_{\scaleto{-2.0}{4pt}}^{\scaleto{+2.0}{4pt}}$ & $-3.88_{\scaleto{-0.38}{4pt}}^{\scaleto{+0.38}{4pt}}$ & $5.59_{\scaleto{-0.21}{4pt}}^{\scaleto{+0.21}{4pt}}$ & \scriptsize{[7,8], RV[9,10]}\\
Carina II & 114.11 & -57.99 & 23 & $36.20_{\scaleto{-0.60}{4pt}}^{\scaleto{+0.60}{4pt}}$ & $17.79_{\scaleto{-0.05}{4pt}}^{\scaleto{+0.05}{4pt}}$ & 0.185 & $0.34_{\scaleto{-0.07}{4pt}}^{\scaleto{+0.07}{4pt}}$ & $170.0_{\scaleto{-9.0}{4pt}}^{\scaleto{+9.0}{4pt}}$ & $-4.50_{\scaleto{-0.10}{4pt}}^{\scaleto{+0.10}{4pt}}$ & $8.69_{\scaleto{-0.75}{4pt}}^{\scaleto{+0.75}{4pt}}$ & \scriptsize{[5],RV[6,11]}\\
Boötes II & 209.5 & 12.86 & 14 & $42.00_{\scaleto{-1.60}{4pt}}^{\scaleto{+1.60}{4pt}}$ & $18.10_{\scaleto{-0.06}{4pt}}^{\scaleto{+0.06}{4pt}}$ & 0.031 & $0.24_{\scaleto{-0.12}{4pt}}^{\scaleto{+0.12}{4pt}}$ & $-70.00_{\scaleto{-27.0}{4pt}}^{\scaleto{+27.0}{4pt}}$ & $-2.94_{\scaleto{-0.74}{4pt}}^{\scaleto{+0.74}{4pt}}$ & $3.05_{\scaleto{-0.45}{4pt}}^{\scaleto{+0.45}{4pt}}$ & \scriptsize{[12,8],RV[13,14]}\\
Boötes III & 209.25 & 26.8 & 16 & $46.50_{\scaleto{-2.00}{4pt}}^{\scaleto{+2.00}{4pt}}$ & $18.35_{\scaleto{-0.01}{4pt}}^{\scaleto{+0.01}{4pt}}$ & 0.021 & $0.50_{\scaleto{-0.00}{4pt}}^{\scaleto{+0.00}{4pt}}$ & $90.00_{\scaleto{-0.00}{4pt}}^{\scaleto{+0.00}{4pt}}$ & $-5.80_{\scaleto{-0.50}{4pt}}^{\scaleto{+0.50}{4pt}}$ & $30.0_{\scaleto{-0.00}{4pt}}^{\scaleto{+0.00}{4pt}}$ & \scriptsize{[15,16],RV[16,17]}\\
Tucana IV & 0.72 & -60.83 & 11 & $47.00_{\scaleto{-4.00}{4pt}}^{\scaleto{+4.00}{4pt}}$ & $18.41_{\scaleto{-0.19}{4pt}}^{\scaleto{+0.19}{4pt}}$ & 0.012 & $0.39_{\scaleto{-0.10}{4pt}}^{\scaleto{+0.07}{4pt}}$ & $27.00_{\scaleto{-8.0}{4pt}}^{\scaleto{+9.0}{4pt}}$ & $-3.00_{\scaleto{-0.40}{4pt}}^{\scaleto{+0.30}{4pt}}$ & $9.30_{\scaleto{-0.90}{4pt}}^{\scaleto{+1.40}{4pt}}$ & \scriptsize{[18],RV[18]} \\
Boötes I & 210.02 & 14.51 & 159 & $66.00_{\scaleto{-3.00}{4pt}}^{\scaleto{+3.00}{4pt}}$ & $19.11_{\scaleto{-0.08}{4pt}}^{\scaleto{+0.08}{4pt}}$ & 0.017 & $0.25_{\scaleto{-0.02}{4pt}}^{\scaleto{+0.02}{4pt}}$ & $7.00_{\scaleto{-3.0}{4pt}}^{\scaleto{+3.0}{4pt}}$ & $-6.02_{\scaleto{-0.25}{4pt}}^{\scaleto{+0.25}{4pt}}$ & $11.26_{\scaleto{-0.27}{4pt}}^{\scaleto{+0.27}{4pt}}$ & \scriptsize{[19,8], RV[20,21,22,23,24,25]} \\
\hline
\end{tabular}}
\end{center}
\label{table:table1}
\footnotetext{}{\footnotesize Note: Properties of the {satellite galaxies} that are subjected to this study are listed in order of increasing distance. {Satellite galaxies} with no RV measurements are not included. Columns: Satellite name, RA, DEC, number of known spectroscopic members, heliocentric distance, distance modulus, reddening, ellipticity, position angle, absolute magnitude, half light radius and associated references. The reddening was computed using the average of individual member star reddening based on values from \citet{Schlegel_1998}. \\Reference. (1) \citet{Drlica-Wagner_2015};(2) \citet{Simon};(3) \citet{Li_2018};(4) \citet{Koposov};(5) \citet{Torrealba};(6) \citet{Li2018b};(7) \citet{Koposov2015};(8) \citet{Muñoz_2018};(9) \citet{Simon2015};(10) \citet{Koposov2015b};(11) \citet{Ji_2020};(12) \citet{Walsh};(13) \citet{Koch};(14) \citet{Bruce_2023};(15) \citet{Grillmair_2009};(16) \citet{Carlin_2009};(17) \citet{Carlin_2018};(18) \citet{Simon_2020};(19) \citet{Dall'Ora2006};(20) \citet{Muñoz_2006};(21) \citet{Martin_2007};(22) \citet{Jenkins_2021};(23) \citet{Longeard_2022};(24) \citet{Brown_2014};(25) \citet{Waller_2022};}
\end{table*}
\section{Methods}
\label{sec-methods}
\subsection{Clustering Algorithms}
Most clustering algorithms work by minimizing the distance of points from the cluster center for a given number of clusters (e.g., K-means; \citet{MacQueen_1967}). These algorithms work quite well for a rounded data distribution; however, they do not perform well in rejecting the background. Gaussian mixtures \citep{MCLA_2000}, on the other hand, has a similar approach as K-means; it can identify overlapping clusters and backgrounds and assign probability. In density-based clustering algorithms, both spatial and spectral domains (Fourier domain) are considered to effectively remove the noise and low-density regions \citep{Ester_1996}. This approach is later extended and applied hierarchically \citep{Campello_2013}. In order to locate clusters of points (in this case, stars), density-based clustering algorithms search for areas in the parameter space of the data that have a high density of points and are enclosed by areas within a lower density. The density-based clustering approach is not sensitive to the shape of the clusters. DBSCAN (Density-Based Spatial Clustering of Applications with Noise; \citet{Ester_1996}) is a  density-based unsupervised clustering algorithm. HDBSCAN (Hierarchical Density-Based Spatial Clustering of Applications with Noise, \citet{Campello_2013}, implemented in Python by \citet{McInnes_2017}) extends DBSCAN by transforming it into a hierarchical clustering method and then extracting a flat clustering, based on cluster stability. As indicated by their names, DBSCAN and HDBSCAN are the exclusive classes of clustering algorithms that can eliminate points (noise) that do not reside within locally dense regions. Although DBSCAN and HDBSCAN have yet to find application within the realm of dwarf galaxy research, there exist analogous investigations in the field of star clusters where these machine learning techniques have been employed\citep{Pasquato_2019, Castro-Ginard_2022}.
\subsubsection{DBSCAN}
DBSCAN works by defining two hyperparameter values: a distance metric, such as the Euclidean distance, the maximum distance between two points in order for them to be considered as part of a given cluster  ({\fontfamily{lmtt}\selectfont eps} in {\fontfamily{lmtt}\selectfont scikit-learn python package}) and a minimum number of points required to assign a cluster status ({\fontfamily{lmtt}\selectfont min\textunderscore samples} in {\fontfamily{lmtt}\selectfont scikit-learn python package}). The algorithm determines three points; core, boundary, and noise points; based on their relationship with other points in the dataset. A core point meets the criteria of having at least a minimum number of points ({\fontfamily{lmtt}\selectfont min\textunderscore samples}) within an {\fontfamily{lmtt}\selectfont eps} distance. Although a border point is not a core point, it will at least have another core point nearby. A noise or outlier point is any point other than a core or border point. Creating clusters involves selecting a core point from the data set, looking for any nearby points, and placing them in the same cluster as the initial core point. This process is repeated until all points are assigned to a cluster (core and border) or deemed an outlier (noise). One of the advantages of DBSCAN is its ability to identify clusters of arbitrary shapes and sizes. It can also handle datasets with varying densities, returning any number of clusters. Although DBSCAN is less sensitive to parameter selection than other clustering methods, it still necessitates careful selection of hyperparameters. Inappropriate parameter selection might cause the data to be over- or under-fitted, leading to erroneous cluster allocations. 
\subsubsection{HDBSCAN}
HDBSCAN is designed to function with an unknown number of groups with varying shapes and density gradients. HDBSCAN calculates the density surrounding each point and builds a hierarchical cluster tree using robust single linkage clustering based on the density information. One can cut at variable heights to decide the smallest sized cluster based on a few hyperparameters, mainly {\fontfamily{lmtt}\selectfont min\textunderscore cluster\textunderscore size} in {\fontfamily{lmtt}\selectfont hdbscan python package\footnote{https://pypi.org/project/hdbscan/}}. These can be thought of as stable clusters (with
a {\fontfamily{lmtt}\selectfont min\textunderscore cluster\textunderscore size}) with less probable members evaporating from
them as they ascend the hierarchical tree. This process assures that
the emerging clusters are relatively stable beyond density thresholds. As a result, it is sensitive to datasets with variable densities of valid groups.
Cutting through the tree may acquire a similar result to executing DBSCAN for a specific value of {\fontfamily{lmtt}\selectfont eps} and {\fontfamily{lmtt}\selectfont min\textunderscore samples}. HDBSCAN determines the most effective path across the dendrogram by returning the most stable clusters. The capability to recognize clusters of different densities makes HDBSCAN superior to DBSCAN. However, DBSCAN may be more appropriate in some instances where the data is known to have a specific structure or where computational efficiency is a concern.
\subsection{Hyperparameters} 
We investigate potential hyperparameter choices for each algorithm to determine their influence on clustering performance. 
\begin{enumerate}
\item Distance metrics:  The best options providing comparable outcomes and performance are {\fontfamily{lmtt}\selectfont euclidean} and {\fontfamily{lmtt}\selectfont manhattan} distance metrics. We decided to employ {\fontfamily{lmtt}\selectfont euclidean} distance for this work.
\item {\fontfamily{lmtt}\selectfont eps}:  To identify the neighbourhood radius for DBSCAN, we used the elbow method, which involves plotting the distances of each spectroscopic member point to its $k^{th}$ nearest neighbor against $k$. We selected the $k^{th}$ distance value where the graph shows an elbow in slope. We use this value as a starting point for {\fontfamily{lmtt}\selectfont eps}.
\item {\fontfamily{lmtt}\selectfont min\textunderscore samples}: The number of points in a neighbourhood required for a point to be identified as a core point is iterated over a search space of values ranging from 2 to 100.  
\item {\fontfamily{lmtt}\selectfont min\textunderscore cluster\textunderscore size}: Due to the possibility of creating an irrational number of clusters, we avoided using smaller values for {\fontfamily{lmtt}\selectfont min\textunderscore cluster\textunderscore size} in HDBSCAN. Two options were explored: 
\begin{enumerate}  
    \item{{\fontfamily{lmtt}\selectfont min\textunderscore cluster\textunderscore size} is iterated over a search space of values ranging from 5 to 200, with or without fixing {\fontfamily{lmtt}\selectfont allow\textunderscore single\textunderscore cluster} = True. Setting "{\fontfamily{lmtt}\selectfont allow\textunderscore single\textunderscore cluster} = True" returns only one prominent cluster in a given field.}
    \item{{\fontfamily{lmtt}\selectfont min\textunderscore cluster\textunderscore size} and {\fontfamily{lmtt}\selectfont min\textunderscore samples} are iterated over a search space \{10,100\} and \{5,50\} respectively, with or without fixing {\fontfamily{lmtt}\selectfont allow\textunderscore single\textunderscore cluster} = True.}
\end{enumerate}
\item The default parameter Excess of Mass ({\fontfamily{lmtt}\selectfont eom}) proved to be the most effective when we investigated the {\fontfamily{lmtt}\selectfont cluster\textunderscore selection\textunderscore method} in HDBSCAN.
\end{enumerate}
\subsubsection{Selection of Hyperparameters}
\begin{figure*}
    \centering
    \includegraphics[width=0.4\textwidth]{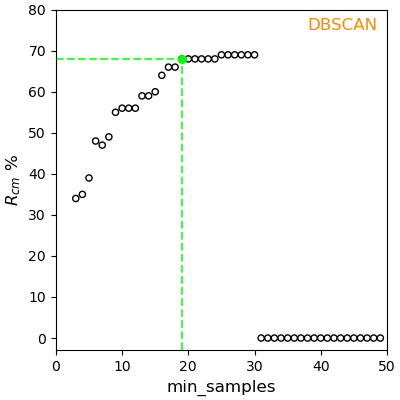}
    \includegraphics[width=0.4\textwidth]{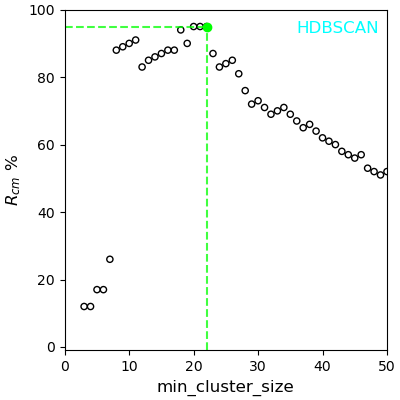}
    \includegraphics[width=0.4\textwidth]{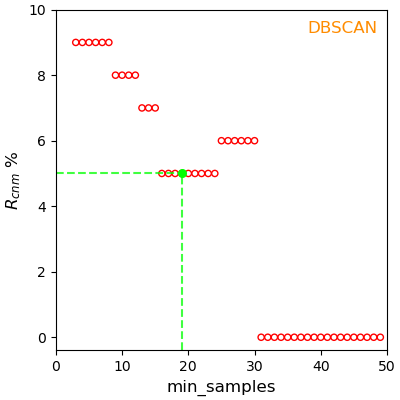}
    \includegraphics[width=0.4\textwidth]{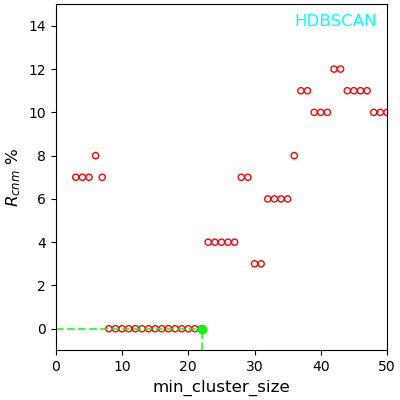} 
    \caption{{\textit{Left - Top}: {\fontfamily{lmtt}\selectfont min\textunderscore samples} vs. candidate recovery rate ($R_{cm}$) for RetII in DBSCAN clustering for an {\fontfamily{lmtt}\selectfont eps} value 0.4. \textit{Left - Bottom}: shows the candidate non-member recovery rate ($R_{cnm}$) vs. {\fontfamily{lmtt}\selectfont min\textunderscore samples}. The \textit{Right - Top \& Bottom} plots show the hyperparameter selection for HDBSCAN. Variation of $R_{cm}$(\textit{Top}) and $R_{cnm}$(\textit{Bottom}) with hyperparameter {\fontfamily{lmtt}\selectfont min\textunderscore cluster\textunderscore size} has been plotted. The green filled circle with dotted lines in all four panels denote finally selected hyperparameter, considering high ($R_{cm}$) and low ($R_{cnm}$).}}
    \label{fig2:val-db-r}
\end{figure*}

\begin{figure*}
    \centering
    \includegraphics[width=0.4\textwidth]{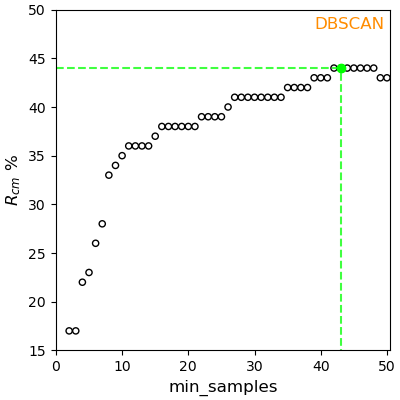}
    \includegraphics[width=0.4\textwidth]{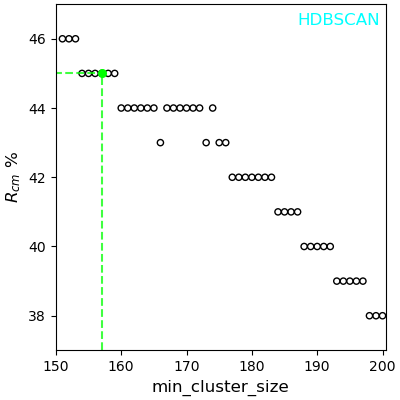}
    \includegraphics[width=0.4\textwidth]{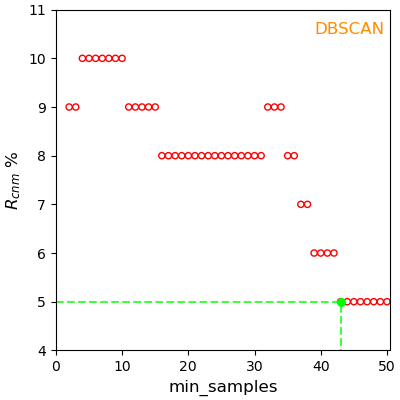}
    \includegraphics[width=0.4\textwidth]{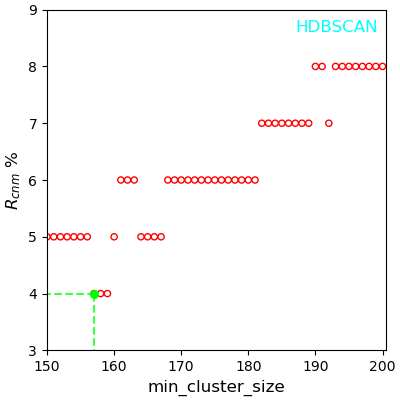} 
    \caption{{Same as Figure \ref{fig2:val-db-r}, but for BoöI. \textit{Left - Top}: Hyperparameter selection showing {\fontfamily{lmtt}\selectfont min\textunderscore samples} vs. candidate recovery rate ($R_{cm}$) in DBSCAN clustering for an {\fontfamily{lmtt}\selectfont eps} value 0.55. \textit{Left - Bottom}: {Variation of $R_{cnm}$ with increasing \fontfamily{lmtt}\selectfont min\textunderscore cluster\textunderscore size} for BoöI in DBSCAN clustering has been shown. \textit{Right - Top \& Bottom} plots show the hyperparameter selection for HDBSCAN. Variation of  $R_{cm}$(\textit{Top}) and $R_{cnm}$(\textit{Bottom}) with hyperparameter {\fontfamily{lmtt}\selectfont min\textunderscore cluster\textunderscore size} has been plotted. The green filled circle with dotted lines in all four panels denote finally selected hyperparameter.}}
    \label{fig3:val-db-b}
\end{figure*}

{
We compute recovery rates for each iteration in which a certain hyperparameter is examined. The candidate member recovery rate is determined as follows:
\begin{equation}
\label{eq:1}
{\scriptstyle \text{$R_{cm}$}\%}  = \dfrac{{\scriptstyle\text{No: of spec. members identified by algorithm}}}{{\scriptstyle \text{Total number of candidates identified by algorithm}}}{\scriptstyle\text{ x 100}}
\end{equation} 
The candidate non-member recovery rate is determined as:
\begin{equation}
\label{eq:2}
{\scriptstyle  \text{$R_{cnm}$}\%}   = \dfrac{{\scriptstyle\text{No: of spec. non-members identified by algorithm}}}{{\scriptstyle \text{Total number of candidates identified by algorithm}}}{\scriptstyle\text{ x 100}}
\end{equation} 
We choose the hyperparameters depending on a high \text{$R_{cm}$}\% and a low  \text{$R_{cnm}$}\%. This has helped to a certain extent in removing the noise points(fore-ground MW stars) from the final candidates. The member recovery rate has been consistently monitored across various parameters and has served as a criterion for selecting hyperparameters in situations where neither $R_{cm}$\% nor $R_{cnm}$\% could guide the hyperparameter selection. The member recovery rate ($R_{m}$\%) is given by: 
\begin{equation}
\label{eq:3}
{\scriptstyle \text{$R_{m}$}\%}  = \dfrac{{\scriptstyle\text{No: of spec. members identified by algorithm}}}{{\scriptstyle \text{Total number of spectroscopic members}}}{\scriptstyle\text{ x 100}}
\end{equation} 
We tried to achieve a minimum threshold of 67\% for the member recovery rate, i.e., the algorithm must recover at least $\frac{2}{3}$ of the spectroscopic members to be deemed valid clustering. In some cases, certain {satellite galaxies} returned high  \text{$R_{cm}$}\% and a low  \text{$R_{cnm}$}\%  with ease (e.g., RetII: refers to Reticulum II; short names of {satellite galaxies} listed in Table \ref{table:summary}). 
On the other hand, a satisfactory recovery seemed difficult for the {satellite galaxies} with elongated structures or tidal tails (e.g., BoöI, TucIII). As an example, we show two {satellite galaxies} with different characteristics to demonstrate the hyperparameter selection. We chose RetII as one of the {satellite galaxies}, due to its proximity and the highest anticipated membership. Additionally, it has a significant number of spectroscopic members, with the clear clustering of stars in proper-motion space and well-constrained satellite parameters. Along with a richly populated RGB and HB, the elongated and structured spatial distribution of its stars and isophote twists (\citet{Fellhauer_2008}, \citet{Roderick_2016}) suggestive of tidal stretching make BoöI an ideal candidate for demonstration. Figure \ref{fig2:val-db-r} shows the variation of candidate member ($R_{cm}$) and non-member recovery rates ($R_{cnm}$) for a range of DBSCAN parameters ({\fontfamily{lmtt}\selectfont eps} vs. {\fontfamily{lmtt}\selectfont min\textunderscore samples}) and HDBSCAN parameter ({\fontfamily{lmtt}\selectfont min\textunderscore cluster\textunderscore size}). The \textit{Top Left} and \textit{Right} panels in Figure \ref{fig2:val-db-r} illustrate the $R_{cm}$ of RetII in DBSCAN and HDBSCAN, respectively. The \textit{Bottom} panels in Figure \ref{fig2:val-db-r} show the $R_{cnm}$ in DBSCAN and HDBSCAN for RetII. Similarly, Figure \ref{fig3:val-db-b} illustrates the variation in $R_{cm}$ and $R_{cnm}$ for BoöI in DBSCAN (\textit{Left}) and HDBSCAN algorithm(\textit{Right}). In Figures \ref{fig2:val-db-r} and \ref{fig3:val-db-b}, the green closed circle with dotted lines gives finally selected hyperparameters in each case. To mitigate erroneous cluster allocations and focus on conventional clusters within a given field, we implemented a cut-off criterion by setting a maximum limit of four for the number of clusters retrieved. This eliminates the presence of multiple stellar systems in the background/foreground of each satellite field (e.g., CarIII in the CarII field, and vice versa).}

\section{Results and Discussions}
\label{sec-results-discussions}
\begin{figure*}
    \centering
    \includegraphics[width=0.8\textwidth]{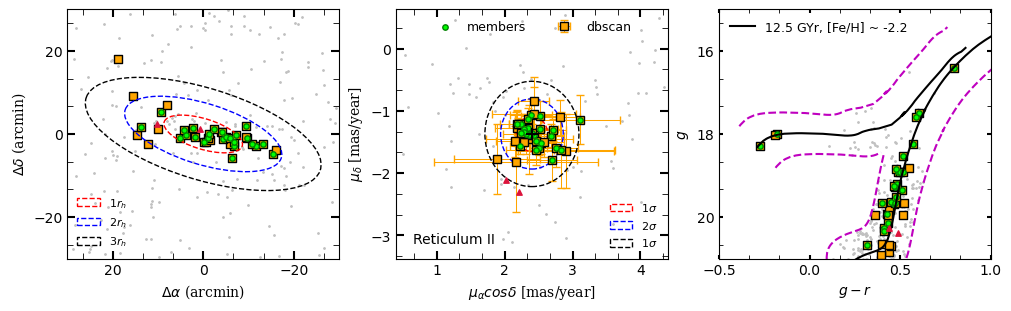} 
    \includegraphics[width=0.8\textwidth]{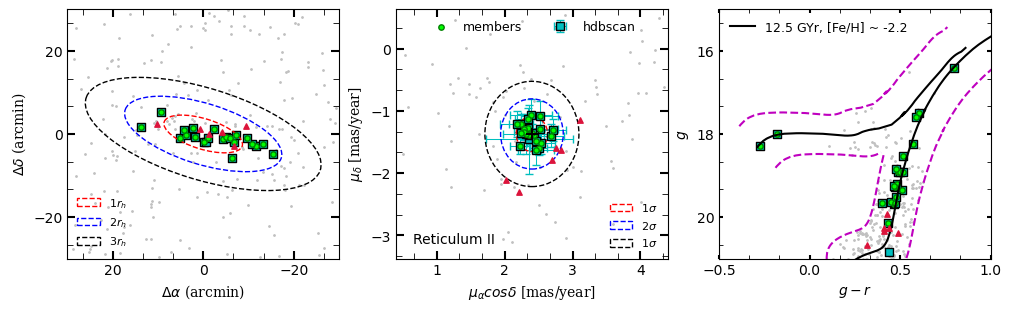} 
    \caption{\textit{Top - Left \& Middle}: The selected candidate members for RetII using the DBSCAN algorithm, with the left (middle) showing the positional (proper-motion) space distribution. Points are colored by: all stars (grey), DBSCAN-selected (orange). Overlaid green circles are the spectroscopically confirmed members (References in Table \ref{table:table1}). In the \textit{Top - Left } panel, the ellipses indicate increasing half-light-radius(red: 1$r_{h}$, blue: 2$r_{h}$, black: 3$r_{h}$). {The dotted ellipses in the \textit{Top - Middle} panel indicate increasing standard deviation of spectroscopic member proper motions (red: 1$\sigma$, blue: 2$\sigma$, black: 3$\sigma$). The black plus sign represents the median proper motion calculated from spectroscopic members employed in this study for each individual galaxy.} \textit{Top - Right}: CMD with BasTi isochrone of age = 12.5 Gyr and [Fe/H] = -2.2 (black). The magenta dotted lines show our color-magnitude selection, which includes a range based on color ($\Delta (g − r)$ = ±0.1 mag) and magnitude ($\Delta g$ = ±0.5 mag) to the isochrone. The \textit{Bottom} panels show similar plots with candidate members (cyan) obtained after HDBSCAN algorithm implementation. Crimson filled triangles show the non-retrieved spectroscopic members (similar meaning in Figure \ref{fig5:hyi1cmd} and Figures \ref{figa1:tuc3cmd} to \ref{figa7:boo1cmd}).}
    \label{fig4:ret2cmd}
\end{figure*}

\begin{figure*}
    \centering
    \includegraphics[width=0.8\textwidth]{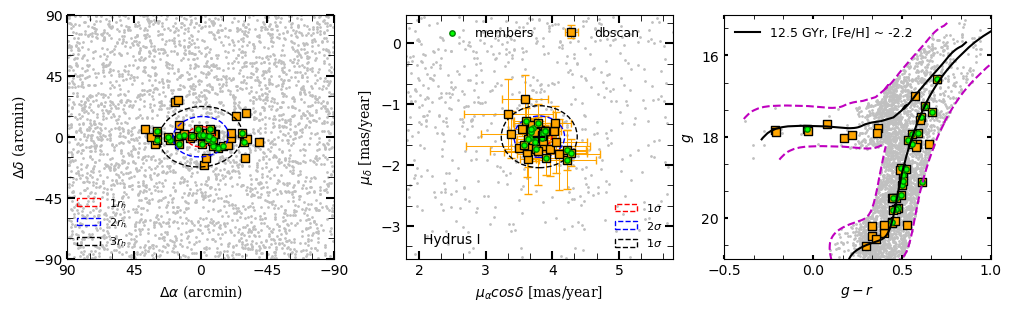} 
    \includegraphics[width=0.8\textwidth]{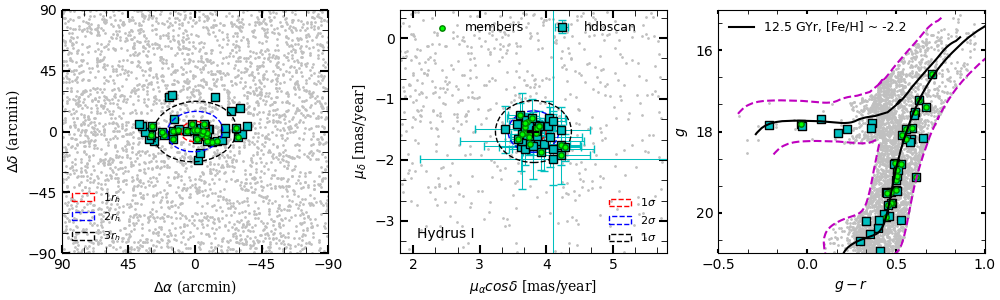} 
    \caption{Same as Figure \ref{fig4:ret2cmd}, but for HyiI. }
    \label{fig5:hyi1cmd}
\end{figure*}

Our results illustrate the efficacy of employing DBSCAN and HDBSCAN as a tool for identifying members of {MW satellite galaxies}. We have achieved a {candidate recovery rate of 50-100\% for all the satellite galaxies in this study (see Table \ref{table:table1}). Both the tools could reject more than 90\% of all the spectroscopic non-members (List of spectroscopic non members for BoöI was kindly provided by Josh Simon, through private communication).}
 {The cluster-identified candidate members showed over-density at the proximity of satellite centers in separate spatial and proper-motion spaces. For example, the upper right panel in Figure \ref{fig4:ret2cmd} illustrates all the cluster-identified targets that outstripped the color-magnitude cut, along with the selected isochrone and color-magnitude selection. The candidate members (orange squares) of RetII selected after implementing the DBSCAN algorithm are shown in the upper left (position space) and middle (proper-motion space) panels of Figure \ref{fig4:ret2cmd}. The green circles are the spectroscopically confirmed members. The dotted ellipses in the upper left (position space) panel are the increasing half-light-radii ($r_{h}$) of RetII from the literature(red: 1$r_{h}$, blue: 2$r_{h}$, black: 3$r_{h}$). The dotted ellipses in the upper middle panel of Figure \ref{fig4:ret2cmd} indicate increasing standard deviation of spectroscopic member proper motions (red: 1$\sigma$, blue: 2$\sigma$, black: 3$\sigma$). The black plus sign represents the median proper motion calculated from spectroscopic members employed in this study for each individual galaxy. The bottom panels of Figure \ref{fig4:ret2cmd} show the candidate members (blue squares) of RetII obtained using HDBSCAN algorithm. Figure \ref{fig5:hyi1cmd} shows similar plots for Hydrus I. In most cases, spectroscopic non-members are two-three times the number of spectroscopic members, which indicates the efficiency of the clustering method for such applications. The non-retrieval of spectroscopic members of certain satellite galaxies could be due to their diluted density in positional and proper-motion space.} 
\subsection{Comparing Clustering Algorithms}
 \par Table \ref{table:summary} presents a summary of validation using the known spectroscopic members and non-members with newly identified member candidates using DBSCAN and HDBSCAN. Table \ref{table:summary} lists the number of common samples from both the methods. The difference is more for larger sparsely populated {satellite galaxies}.
The performance of a density-based clustering method may benefit from the data covering a large volume of the sky, especially for massive {satellite galaxies}, to cover an extensive range of spatial frequencies. In some sparsely populated disrupted {satellite galaxies}, the spectroscopic member retrieval is less than the smaller compact galaxies (in spatial and PM space), possibly due to the restriction of the field. \par
\begin{figure}[ht!]
    \centering
    \includegraphics[width=0.46\textwidth]{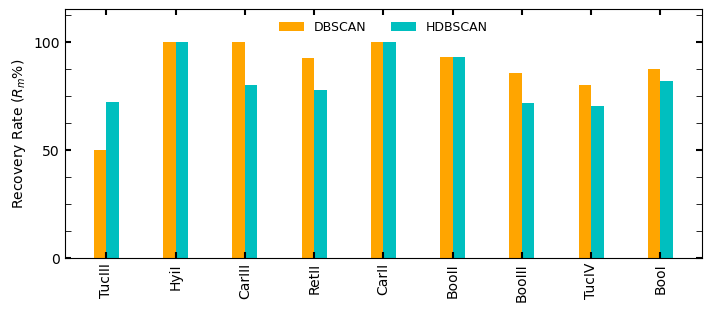} 
    \includegraphics[width=0.46\textwidth]{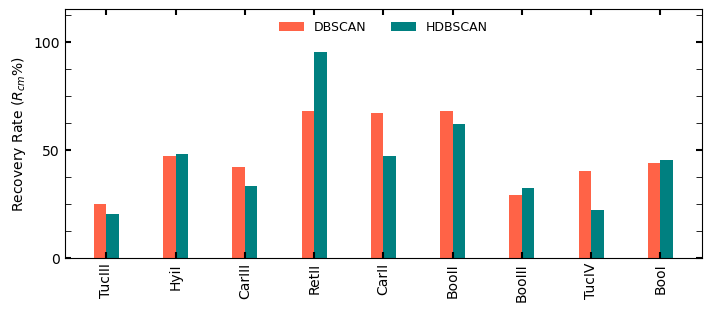} 
    \includegraphics[width=0.46\textwidth]{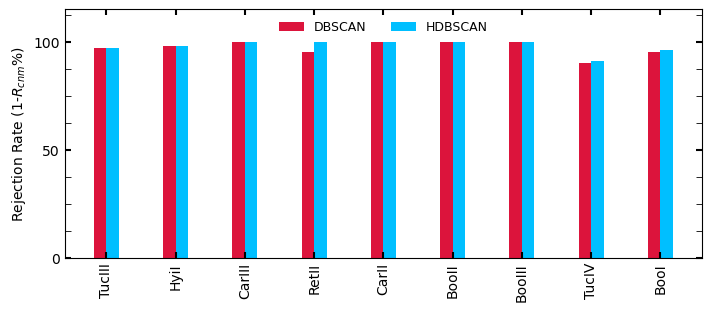}
    \caption{\textit{Top} panel shows the comparison between DBSCAN (orange) and HDBSCAN (cyan) for each satellite in terms of spectroscopic member recovery rate. \textit{Middle} panel  gives the  candidate recovery rate in DBSCAN (coral) and HDBSCAN (teal). \textit{Bottom} panel gives the spectroscopic non-member rejection rates using  DBSCAN (crimson) and HDBSCAN (skyblue). Both the algorithms rejected all the spectroscopic non-members in CarIII, Car II, BoöII and BoöIII.}
    \label{fig7:comp}
\end{figure}

 The number of spectroscopic members retrieved by TucIII,  BoöIII, TucIV and BoöI are relatively lesser (less than 85\%) compared to other satellite galaxies.
In the case of TucIII, the core and tidal tails of TucIII contain a total of 48 spectroscopic members (36 present in Gaia - DECam catalogs). However, the implementation of DBSCAN (HDBSCAN) has successfully identified only 18 (26) of these members. {TucIII has extended tidal tails, and we find members extending till the edge of the field that is considered using HDBSCAN}. A larger area may yield better results through better characterization of low-density spatial frequencies. {However, when utilizing DBSCAN, it is evident that the algorithm still needs to capture stars within the tidal tails. Additionally, our experiments with varying hyperparameters in DBSCAN ({\fontfamily{lmtt}\selectfont eps} and {\fontfamily{lmtt}\selectfont min\textunderscore samples}) resulted in a significant number of non-members, likely due to the density of TucIII being lower than the average density of the MW \citep{Pace_2022}.
The elevated velocity dispersion, small pericenter, and the potential association with the Styx stream (agreement with the predicted retrograde motion of the coincident Styx stellar stream; \cite{Carlin_2018}), BoöIII has been suggested to be experiencing tidal disruption \citep{Carlin_2009}. The lower member recovery rates in BoöIII could be attributed to its relatively low average density \citep{Carlin_2018}, pointing to its status as a potential tidally disrupting galaxy candidate.}
One of the reasons for lower candidate member recovery rates in TucIV can be the sparsely spaced members in spatial and PM space. The scattered structure of TucIV might have been caused by a collision with the Large Magellanic Cloud, which may have altered its trajectory and internal kinematics \citep{Sales_2017,Simon_2020}. Additionally, one of the spectroscopic members in TucIV was later found to be an outlier in the proper motion ($\mu_{\alpha} cos\delta$ = 0.900 [mas/year], $\mu_\delta$= -0.225 [mas/year]).This star is positioned significantly away from the proper motion of TucIV ($\mu_{\alpha} cos\delta$ = 0.534 ± 0.050, $\mu_\delta$ = 1.707 ± 0.055), and is also reported in \citet{Pace_2022}. {The lower rate of recovering members in TucIV might be attributed to the overlap in proper motion between TucIV and foreground stars in the MW \citep{Pace_2019} and the spread in PM space. In TucIV, there is a slight shift observed in the center of the galaxy compared to the results in \citet{Drlica-Wagner_2015} in both DBSCAN and HDBSCAN. As a result, we have opted to utilize the revised center coordinates provided in \citep{Simon_2020}. Therefore, even though this work does not include a revised estimation of structural parameters, this approach has the potential to estimate the density profiles and spatial morphology of {ultra-faint dwarf galaxies}; unlike K-means, satellite center and assumption of the shape of the satellite are not inputs to the methods.}
In the case of BoöI, out of 147 spectroscopic members (87 present in Gaia-DELVE catalogs), the DBSCAN (HDBSCAN) approach successfully retrieves only 76 (71) spectroscopic members. The procedures also have not recovered the five proper-motion non-members reported in \citet{Simon_2018} and the members which are spread over large area in spatial and proper-motion space. 
\renewcommand{\arraystretch}{1.2}
\setlength{\tabcolsep}{3.5pt}
 \begin{table*}
\caption{Candidates identified from clustering algorithms}
\label{table:summary}
\begin{center}
{\footnotesize
\begin{tabular}{|cccccccccccc|}
\hline
Satellite&$\text{N}_{\text{gaia-decam}}$&$\text{N}_{\text{n-gaia-decam}}$&$\text{N}_{\text{d-reco}}$&$\text{N}_{\text{d-nrej}}$&$\text{N}_{\text{d-new}}$&$\text{N}_{\text{hd-reco}}$&$\text{N}_{\text{hd-nrej}}$&$\text{N}_{\text{hd-new}}$&$\text{N}_{\text{com}}$&$\text{N}_{\text{\tiny PEL22}}$&$\text{N}_{{\text{\tiny BTTF22}}}$\\
\hline
Tucana III (TucIII) & 36 &215 & 18 &213& 53 & 26 &211& 99 & 47&52,45&38,28\\
Hydrus I (HyiI) & 22 & 54 & 22 &53& 24 & 22 &53& 23 & 43&43,43&46,46\\
Carina III (CarIII) & 5 & 113 & 5 &113& 7 & 4 &113& 8 & 10&7,6&8,7\\
Reticulum II (RetII) & 27 & 19 & 25 & 17 & 10 & 21 &19& 1 & 22&36,21&37,22\\
Carina II (CarII) & 22 & 154& 22 &154& 11& 22& 154 &25&33&33,42&33,44\\
Boötes II (BoöII) & 14 & 5 & 13 &5& 6 & 13 &5& 8 & 19&17,17&16,17\\
Boötes III (BoöIII) & 14 &3& 13 &3& 34 & 10 &3&21 & 31&44,28&44,28\\
Tucana IV (TucIV) & 10 &48 & 8 &46& 10 & 8 &43& 22& 19&14,14&12,12\\
Boötes I (BoöIII) & 87& 104  & 76 &95& 88 & 71 &97& 74 & 154&162,148&162,149\\
\hline
\end{tabular}}
\end{center}
\footnotetext{}{\footnotesize Note: Column 1 is the Satellite name with short name in brackets. {$\text{N}_{\text{gaia-decam}}$ is the number of spectroscopic members in which Gaia four-paramter astrometric and DeCAM photometry data is available.$\text{N}_{\text{n-gaia-decam}}$ is the number of spectroscopic non-members in which Gaia four-paramter astrometric and DeCAM photometry data is available.} $\text{N}_{\text{d-reco}}$ is the number of spectroscopic members recovered using DBSCAN algorithm. {$\text{N}_{\text{d-nrej}}$ is the number of spectroscopic non-members correctly rejected as outliers using DBSCAN algorithm.} $\text{N}_{\text{d-new}}$ is the new candidate members using DBSCAN. $\text{N}_{\text{hd-reco}}$ is the number of spectroscopic members recovered using HDBSCAN algorithm. {$\text{N}_{\text{hd-nrej}}$ is the number of spectroscopic non-members correctly rejected as outliers using HDBSCAN algorithm. }$\text{N}_{\text{hd-new}}$ is the new candidate members using HDBSCAN. $\text{N}_{\text{com}}$ is the total number of common candidates newly identified using DBSCAN and HDBSCAN.{$\text{N}_{\text{PEL22}}$ is the number of candidates present in \citep{Pace_2022} with membership probability $P_{i} > 0.1$ using DBSCAN (HDBSCAN) algorithm. $\text{N}_{\text{BTTF22}}$ is the number of candidates present in \citep{Battaglia_2022} with membership probability $P_{i} > 0.1$ using DBSCAN (HDBSCAN) algorithm.}}
\end{table*}
{DBSCAN and HDBSCAN exhibited varying behaviors in each satellite, depending on their spatial structure, extended area, and clustering in proper-motion space. 
Figure \ref{fig7:comp} compares DBSCAN and HDBSCAN concerning the retrieval of spectroscopic members and the rejection of non-members in each satellite. It is evident from Figure \ref{fig7:comp} that DBSCAN performed relatively well in recovering spectroscopic members. This can be due to our approximate prior knowledge of the parameter {\fontfamily{lmtt}\selectfont eps} from the elbow method (using spectroscopic members), which returns a cluster with a specific structure and recovery efficiency. All the galaxies except TucIII, CarIII, BoöIII, TucIV and BoöI showed a candidate recovery rate $\sim 50\%$ (Figure \ref{fig7:comp} \textit{Middle} panel). As discussed in Section \ref{sec-results-discussions}, a key factor contributing to the reduced candidate member recovery rates in these satellite galaxies could be the sparsely spaced members in spatial and PM space (e.g. TucIII, CarIII, BoöI). The other reasons include the similarity in MW background proper motion (TucIV), dwarf density lower than MW average density (TucIII, BoöIII). The non-member rejection rate (Figure \ref{fig7:comp} \textit{Bottom} panel)is comparatively similar for both DBSCAN and HDBSCAN. Both the algorithms rejected all the spectroscopic non-members in CarIII, Car II, BoöII, and BoöIII.} 

\subsection{Comparison with the literature}
\begin{figure}[ht!]
    \centering
    \includegraphics[width=0.49\textwidth]{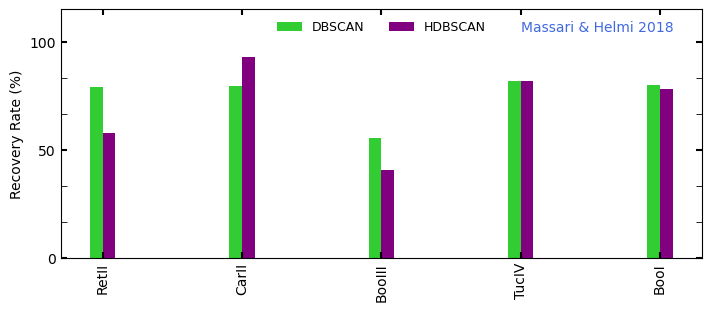} 
    \caption{\textit{Top}: The recovery rate for the {satellite galaxies} common in this work and \citet{Gaia-Collaboration2018}\/\citet{Massari-Helmi_2018} for DBSCAN (light green) and HDBSCAN (purple). The recovery rate in this context refers to the ratio of the number of cluster selected candidates present in \citet{Gaia-Collaboration2018} and \citet{Massari-Helmi_2018} catalogs to the total cluster identified candidates in DBSCAN and HDBSCAN.}
    \label{fig8:comp-h}
\end{figure}

Various approaches have been employed in earlier studies to establish a satellite's systemic proper-motion and subsequently determine the likelihood of each star's membership within a specific field of the satellite. Both \citet{Gaia-Collaboration2018} (hereafter GC18) and \citet{Massari-Helmi_2018}(hereafter MH18) employed Gaia astrometry and Gaia color-magnitude diagrams (CMDs) for their selection process, in contrast to the DECam color-magnitude selection used in this work. They utilized a sigma-clipping routine iteratively to determine the proper-motion of each satellite and the likelihood of individual stars being their members. Figure \ref{fig8:comp-h} shows the catalog recovery rate ($\text{R}_{\text{cat}}$) for the {satellite galaxies} common in GC18 (BoöI), MH18 (RetII, CarII, BoöIII, TucIV) and this work. The recovery rate in this context refers to the ratio of the number of cluster-identified candidates (using DBSCAN (light green) or HDBSCAN (purple)) present in GC18 and MH18 catalogs to the total number of stars commonly present in our input sample and GC18 \& MH18 catalogs as given in Equation \ref{eq:2}. 
\begin{equation}
\label{eq:2}
{\scriptstyle\text{R}_{\text{cat}}\% } = \dfrac{{\scriptstyle \text{Number of cluster candidates in catalog with a $P_{i}$}}}{{\scriptstyle\text{Total number of cross-matched sample with a $P_{i}$}}}{\scriptstyle\text{x100}}
\end{equation} 

All the {satellite galaxies} commonly present in our study and GC18 \& MH18 have an average of 50\% or more recovery rate. {BoöIII showed the least recovery rates: ($\sim$50\%) in both DBSCAN and HDBSCAN.} {HDBSCAN obtained a lower recovery rate (\text{$R_{cat}$} $<$ 50\%) than DBSCAN ($R_{cat}$ $\sim$  50\%) in RetII and BoöIII. This could be connected to the lower recovery rate obtained by HDBSCAN relative to DBSCAN in RetII and BoöIII as seen in Figure \ref{fig7:comp} \textit{Middle} panel. Even though CarII could recover 100\% of spectroscopic members using DBSCAN and HDBSCAN, HDBSCAN has a higher recovery rate than CarII compared to DBSCAN.  This could be because HDBSCAN can determine a suitable clustering arrangement through a hierarchical cluster tree without relying on a fixed neighbourhood distance.}

\begin{figure}[hbt!]
    \centering
    \includegraphics[width=0.48\textwidth]{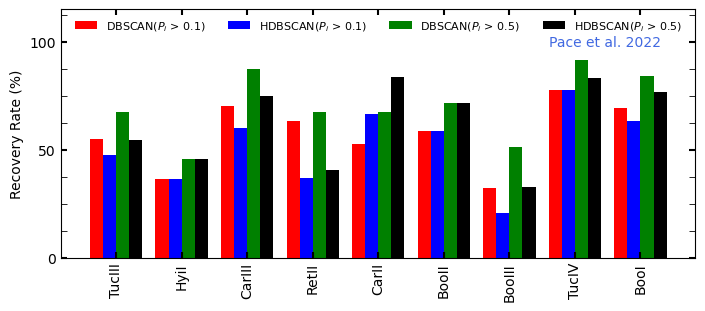} 
    \includegraphics[width=0.48\textwidth]{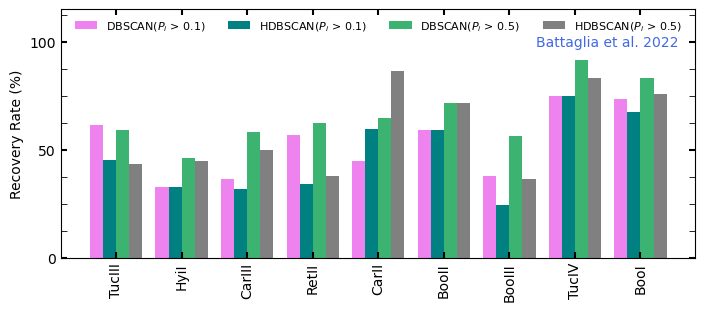} 
    \caption{{\textit{Top:} The recovery rate of candidate members for the satellite galaxies common in this work and \citet{Pace_2022} for DBSCAN (red) and HDBSCAN (blue) with membership probability, $P_{i}$ > 0.1. Green (DBSCAN) and black (HDBSCAN) bars are recovery rate the cluster identified candidates with membership probability, $P_{i}$ > 0.5. \textit{Bottom:} The recovery rate of candidate members in comparison with \citet{Battaglia_2022} for DBSCAN (pink) and HDBSCAN (teal) with membership probability, $P_{i}$ > 0.1. Light green (DBSCAN) and grey (HDBSCAN) bars are recovery rate of the cluster identified candidates with membership probability, $P_{i}$ > 0.5.}}
    \label{fig9:comp-pb}
\end{figure}

The very recent work \citet{Pace_2022} (hereafter PEL22) utilized a gaussian probabilistic mixture models to separate the MW foreground contamination from the satellite stars and determine membership of stars in a satellite, $P_{i}$. PEL22 reports the membership probability ($P_{i}$, for $i^{th}$ star) derived from the  relative likelihood between the satellite and total MW likelihood.  The \textit{Top} panel in Figure \ref{fig9:comp-pb} shows the performance of our method with respect to PEL22 catalog, i.e. a comparison of number of candidate members recovered from PEL22 catalog using DBSCAN (red) and HDBSCAN (blue) with complete membership probability, $P_{i}$ > 0.1.  Green (DBSCAN) and black (HDBSCAN) bars are recovery rate obtained for the cluster identified candidates present in PEL22 catalog which have a membership probability, $P_{i}$ > 0.5. For TucIII, CarIII, RetII (DBSCAN), CarII, BoöII, TucIV and BoöI, our method retrieved about 50\% and more of the candidate members present in PEL22 catalog. The rest of the {satellite galaxies}, HyiI, RetII (HDBSCAN) and BoöIII show lower recovery rates. Despite successfully recovering over 90\% of the spectroscopy members for HyiI and Car II, the recovery rate for PEL22 is considerably low. This variation may be due to the use of NOIRLab Source Catalog (NSC) DR2 (\citet{Nidever_2021}) and Gaia DR3 in PEL22 for HyiI and CarII, while we utilized DELVE DR2 survey. There is a clear offset in color between the CMD selection employed by PEL22 and our selection. Due to the absence of magnitude information in the DELVE DR2 catalog for three spectroscopic members in HyiI, these stars were excluded from the final selection of candidate members. Compared to those with $P_{i}$ > 0.1, satellite candidate members with $P_{i}$ > 0.5 have a higher recovery rate ranging from 10\% to 20\%. 

\citet{Battaglia_2022} (hereafter BTTF22) utilized an adaptable Bayesian approach, as outlined by \citet{McConnachie_2020}, which accounts for star positions, their placement on the color-magnitude map, and their proper-motion characteristics. The \textit{Bottom} panel in Figure \ref{fig9:comp-pb} shows the recovery rate of candidate members in comparison with \citet{Battaglia_2022} for DBSCAN (pink) and HDBSCAN (teal) with membership probability, $P_{i}$ > 0.1. Light green (DBSCAN) and grey (HDBSCAN) bars are recovery rate of the cluster identified candidates with membership probability, $P_{i}$ > 0.5. The number of candidates recovered from PEL22 and BFFT22 catalogs using DBSCAN and HDBSCAN have been tabulated in Table \ref{table:summary}. {From Figure \ref{fig9:comp-pb} it is evident that the PEL22 and BFFT22 have comparable recovery rates from our study. The slight variation in recovery rates could be attributed to the utilization of distinct methodologies for identifying new members and the discrepancy in photometric data sources between PEL22 (DECam) and BFFT22 (Gaia Early-DR3).}

\par {Figure \ref{figa8:comp-pm} compares the median systematic proper motion of the candidate members from the algorithm with the previous known studies. The results from the DBSCAN algorithm demonstrate a stronger level of agreement with the previous studies compared to the results from HDBSCAN. Notably, TucIII and TucIV exhibit more pronounced deviations in the HDBSCAN results. This disparity could be attributed to HDBSCAN identifying an uncertain number of candidates due to the similarity in proper motion between TucIV and foreground stars in the MW and the lower density of TucIII than the overall MW average density. Increasing the field of view (FOV) of the input data by at least doubling its size may enhance HDBSCAN's performance for satellite galaxies with extensive structures or tidal tails.}

\section{Conclusions}
\label{sec-conclusion}
In this paper, we {explored two density-based clustering algorithms (DBSCAN and HDBSCAN) to identify new members of MW satellite galaxies. The data uses accurate astrometry from Gaia DR3 and  DES DR2 \& DELVE DR2 photometry($g,r$). Successful recovery of 75-100\% of spectroscopic members and 95\% rejection of spectroscopic non-members allows the possibility of the density-based unsupervised machine learning algorithms DBSCAN and HDBSCAN for MW {satellite galaxies} and possibly MW streams which do not have symmetric circular or elliptical shape. A large field of view a few times the satellite size might prove the robustness of this method for all {satellite galaxies}, small and large, compact and disrupted irrespective of their shape.} Both the algorithms work relatively better for {satellite galaxies} with more compact density profiles(eg: RetII, HyiI). This approach can be extended to other satellite galaxies and potentially lead to discovering new members of these systems(in preparation).
\section{Acknowledgements}
 We thank Josh Simon and Tom Brown for sharing the relevant data which includes satellite membership from their spectroscopic study. This work has made use of public archival data from European Space Agency (ESA) mission Gaia DR3 (http://www.cosmos.esa.int/gaia) and the Dark Energy
Survey (DES: https://www.darkenergysurvey.org/). The python packages used in this project are {\fontfamily{lmtt}\selectfont astropy} \citep{Astropy-Collaboration_2013}, {\fontfamily{lmtt}\selectfont scikit-learn} (https://scikit-learn.org) and {\fontfamily{lmtt}\selectfont hdbscan} (https://pypi.org/project/hdbscan/). This research has also made use of VizieR catalogue access tool, CDS, NASA's Astrophysics Data System Bibliographic Services and TOPCAT software18 (\citep{Taylor_2011}).
\setlength{\bibsep}{4pt}
\bibliography{jaa_ml}{}

\appendix 
\section*{Appendix A: Selected hyperparameters} 
\renewcommand{\arraystretch}{1.5}
\setlength{\tabcolsep}{4pt}
 \begin{table}[ht!]
\caption{Optimized hyperparameters for DBSCAN and HDBSCAN}
\label{table:hyperparams}
\begin{center}
{\footnotesize
\begin{tabular}{|c|cc|c|}
\hline
Satellite&eps&min\textunderscore samples&min\textunderscore cluster\textunderscore size\\
\hline
TucIII & 0.65 &20 & 24\\
HyiI & 0.45 & 19 & 46\\
CarIII & 0.54 & 10 & 10 \\
RetII & 0.46 & 16 & 22 \\
CarII & 0.35 & 19 & 28 \\
BoöII & 0.59 & 12 & 15 \\
BoöIII & 0.54 &21& 31 \\
TucIV & 0.61 &9 & 32 \\
BoöI & 0.55& 43  & 157 \\
\hline
\end{tabular}}
\end{center}
\footnotetext{}{\footnotesize Note: First column is the Satellite short name. Second column gives the DBSCAN hyperparameters ({\fontfamily{lmtt}\selectfont eps} and {\fontfamily{lmtt}\selectfont min\textunderscore samples}). Third column gives the HDBSCAN hyperparameter ({\fontfamily{lmtt}\selectfont min\textunderscore cluster\textunderscore size}).}
\end{table}
\section*{Appendix B Resultant Plots} 
\begin{figure*}[ht!]
    \centering
    \includegraphics[width=0.9\textwidth]{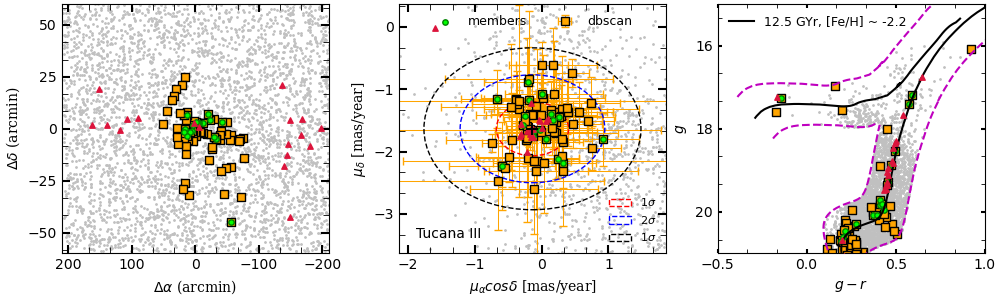} 
    \includegraphics[width=0.9\textwidth]{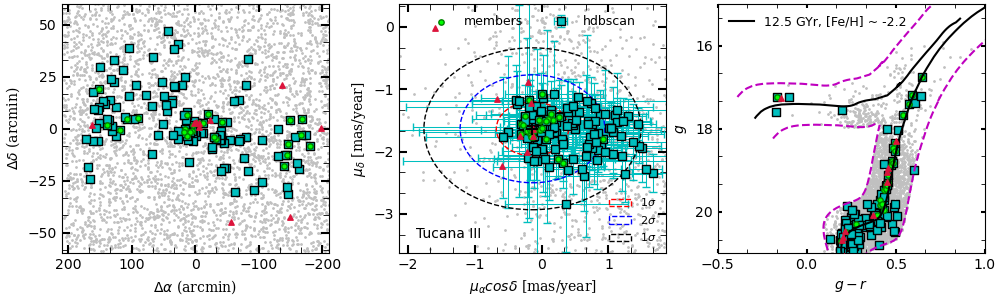} 
    \caption{Same as Figure \ref{fig4:ret2cmd}, but for TucIII.}
    \label{figa1:tuc3cmd}
\end{figure*}

\begin{figure*}[ht!]
    \centering
    \includegraphics[width=0.9\textwidth]{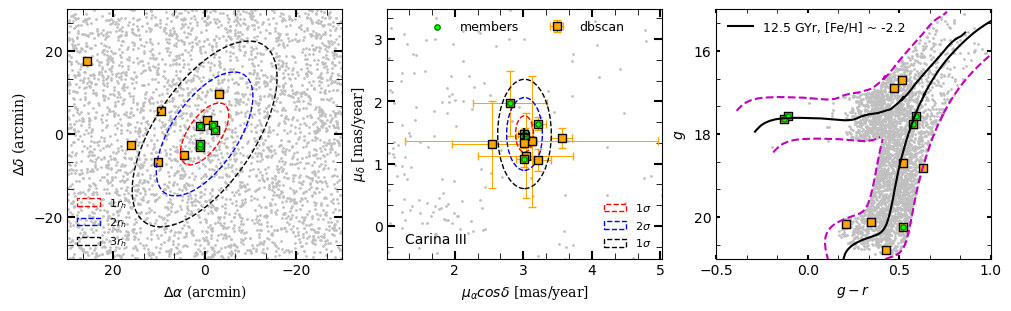} 
    \includegraphics[width=0.9\textwidth]{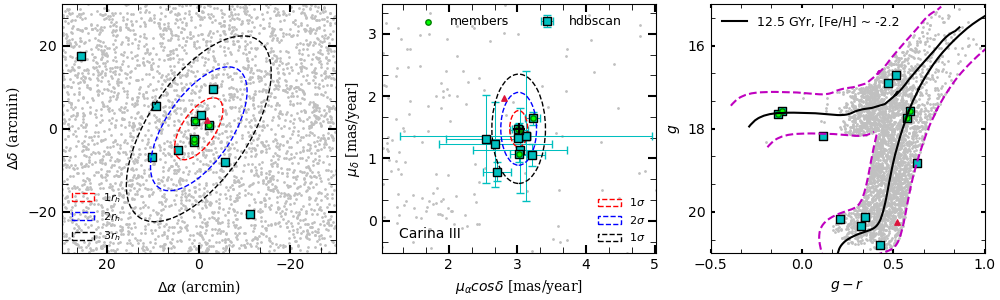} 
    \caption{Same as Figure \ref{fig4:ret2cmd}, but for CarIII.}
    \label{figa2:car3cmd}
\end{figure*}

\begin{figure*}[ht!]
    \centering
    \includegraphics[width=0.9\textwidth]{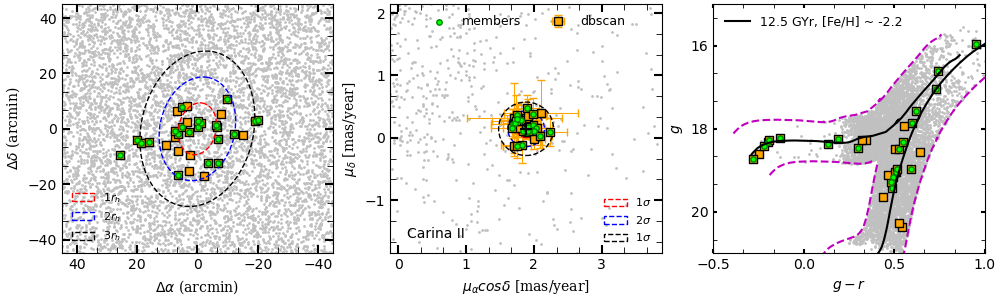} 
    \includegraphics[width=0.9\textwidth]{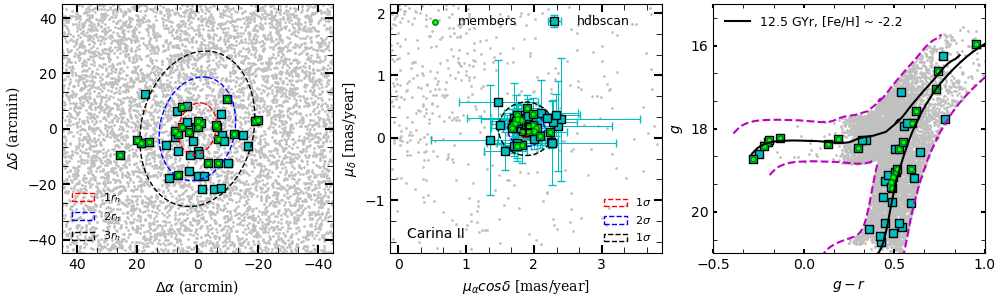} 
    \caption{Same as Figure \ref{fig4:ret2cmd}, but for CarII.}
    \label{figa3:car2cmd}
\end{figure*}

\begin{figure*}[ht!]
    \centering
    \includegraphics[width=0.9\textwidth]{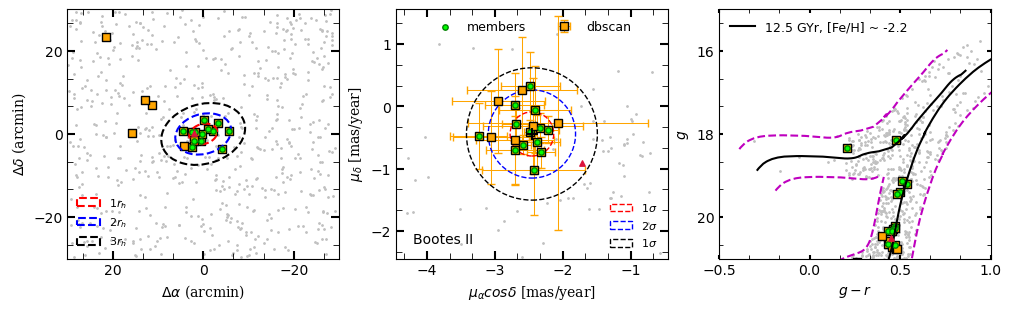} 
    \includegraphics[width=0.9\textwidth]{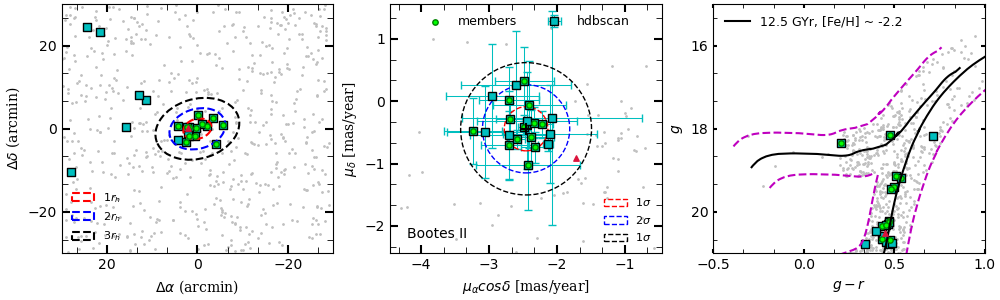} 
    \caption{Same as Figure \ref{fig4:ret2cmd}, but for BoöII.}
    \label{figa4:boo2cmd}
\end{figure*}

\begin{figure*}[ht!]
    \centering
    \includegraphics[width=0.9\textwidth]{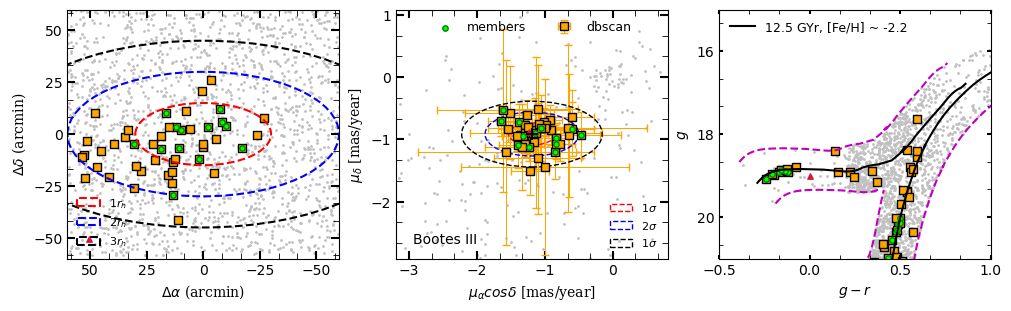} 
    \includegraphics[width=0.9\textwidth]{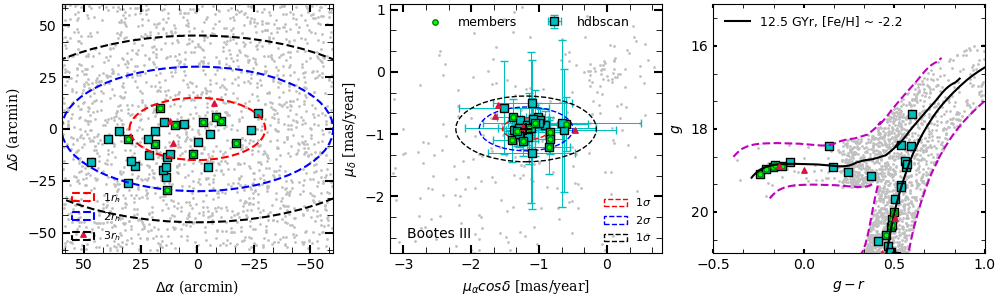} 
    \caption{Same as Figure \ref{fig4:ret2cmd}, but for BoöIII.}
    \label{figa5:boo3cmd}
\end{figure*}

\begin{figure*}[ht!]
    \centering
    \includegraphics[width=0.9\textwidth]{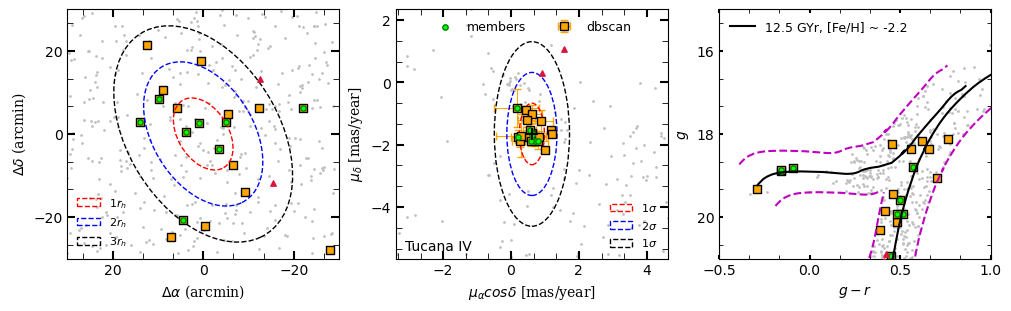} 
    \includegraphics[width=0.9\textwidth]{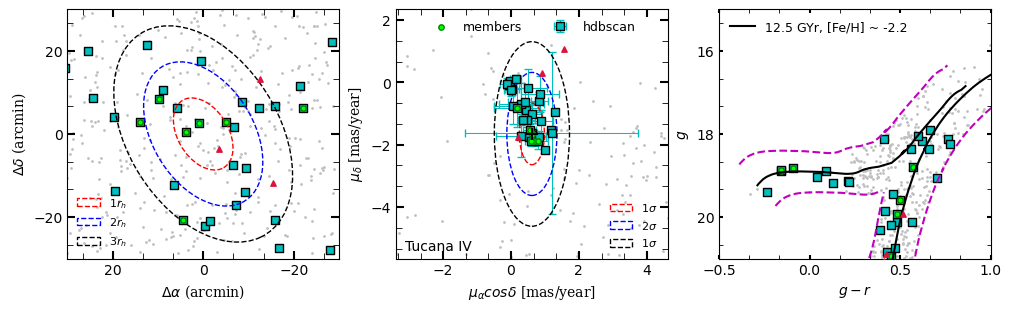} 
    \caption{Same as Figure \ref{fig4:ret2cmd}, but for TucIV.}
    \label{figa6:tuc4cmd}
\end{figure*}

\begin{figure*}[ht!]
    \centering
    \includegraphics[width=0.9\textwidth]{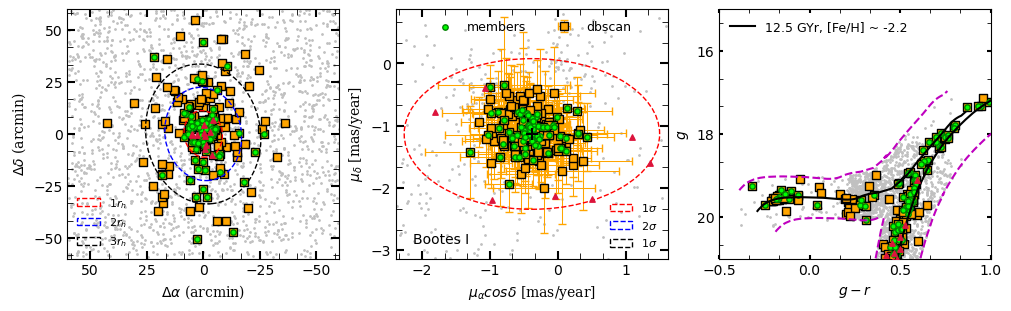} 
    \includegraphics[width=0.9\textwidth]{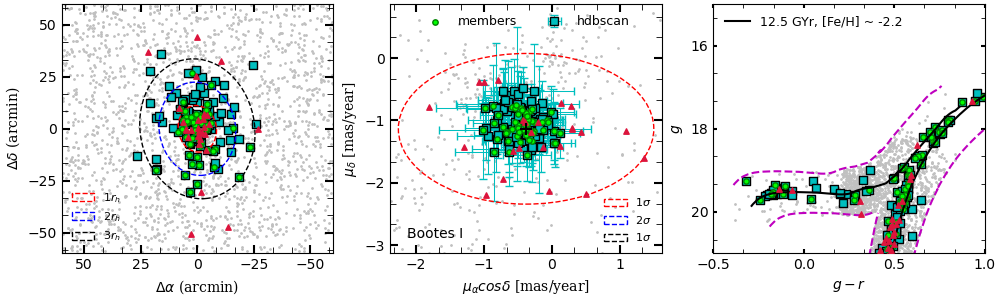} 
    \caption{Same as Figure \ref{fig4:ret2cmd}, but for BoöI.}
    \label{figa7:boo1cmd}
\end{figure*}

\begin{figure*}[ht!]
    \centering
    \includegraphics[width=0.8\textwidth]{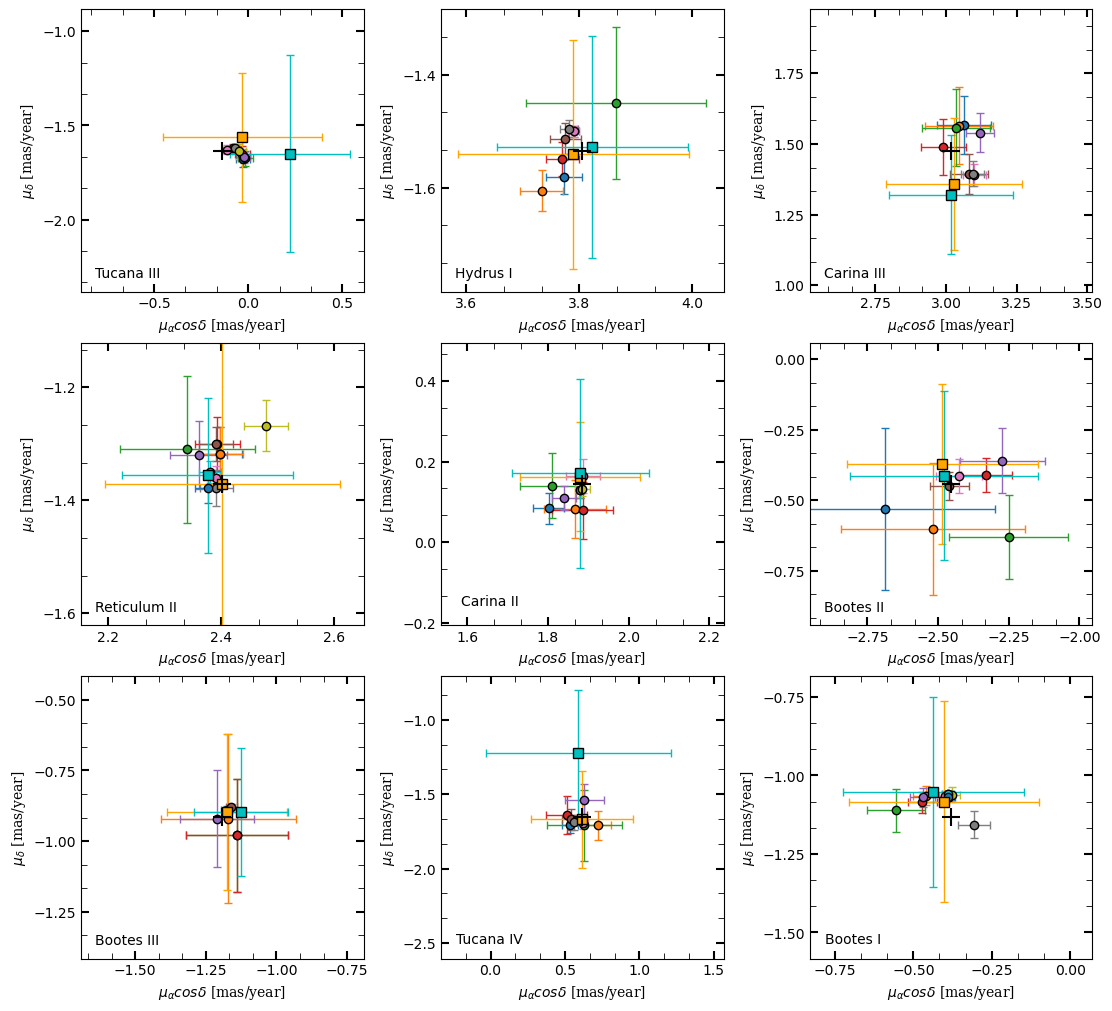} 
    \caption{Comparison of median Proper motion of candidate members from DBSCAN(orange square) and HDBSCAN(cyan square) with other literature values(filled circles with different colors). Error bar shows the standard deviation derived from the proper motion of candidate members.  \\Reference. (1) \citet{Kallivayalil_2018};(2) \citet{Carlin_2018};(3) \citet{Massari-Helmi_2018};(4) \citet{Gaia-Collaboration_2018};(5) \citet{Fritz_2018};(6) \citet{Simon_2018};(7) \citet{Pace_2019};(8) \citet{Simon_2020};(9) \citet{McConnachie_2020};(10) \citet{McConnachie_2020b};(11) \citet{Vitral_2021};(12) \citet{Li_2021};(13) \citet{Bruce_2023};(14) \citet{Martinez-Garcia_2021};(15) \citet{Longeard_2022};(16) \citet{Battaglia_2022};(17) \citet{Pace_2022}}
    \label{figa8:comp-pm}
\end{figure*}

The resultant plots for UFDs using DBSCAN and HDBSCAN algorithms are shown below (Figures \ref{figa1:tuc3cmd} to \ref{figa7:boo1cmd}). Figure \ref{figa8:comp-pm} shows the calculated proper motion from the identified candidates of each UFD using DBSCAN and HDBSCAN has been plotted in comparison with the literature values. 
\newpage

\end{document}